\documentclass[12pt,twoside]{article}
\usepackage{fleqn,espcrc1}

\usepackage[figuresright]{rotating}

\def\Journal#1#2#3#4{{#1} {\bf #2}, #3 (#4)}

\def\NPB{{\em Nucl. Phys.} B}
\def\PLB{{\em Phys. Lett.}  B}

\def\EPC{{\em Euro. Phys.} C}
\def\RPP{\em Rep. Prog. Phys.}

\def\be{\begin{equation}}
\def\ee{\end{equation}}
\def\bea{\begin{eqnarray}}
\def\eea{\end{eqnarray}}

\newcommand{\kk}{\mbox{$K^+K^-$ }}

\newcommand{\pipi}{\mbox{$\pi^+\pi^-$ }}

\newcommand{\AmS}{{\protect\the\textfont2
  A\kern-.1667em\lower.5ex\hbox{M}\kern-.125emS}}

\title{New effects observed in central production by experiment WA102
at the CERN Omega Spectrometer }

\author{A. Kirk\address{School of Physics, Birmingham University,
         Birmingham, U.K.}}

\begin{document}

\maketitle

\begin{abstract}
A partial wave analysis of the centrally produced $K \overline K$ and $\pi \pi$
systems shows that the $f_J(1710)$ has J~=~0.
In addition,
a study of central meson production as a function of the
difference in transverse momentum ($dP_T$)
of the exchanged particles
shows that undisputed $q \overline q$ mesons
are suppressed at small $dP_T$ whereas the glueball candidates
are enhanced and that
the production cross section for different
resonances depends strongly on the azimuthal angle between the
two outgoing protons.
\end{abstract}

\section{INTRODUCTION}
\par
There is considerable current interest in trying to isolate the lightest
glueball.
Several experiments have been performed using glue-rich
production mechanisms.
One such mechanism is Double Pomeron Exchange (DPE) where the Pomeron
is thought to be a multi-gluonic object.
Consequently it has been
anticipated that production of
glueballs may be especially favoured in this process~\cite{closerev}.
\par
The WA102 experiment at the CERN Omega Spectrometer
studies centrally produced exclusive final states
formed in the reaction
\noindent
\begin{equation}
pp \longrightarrow p_{f} X^{0} p_s,
\label{eq:1}
\end{equation}
where the subscripts $f$ and $s$ refer to the fastest and slowest
particles in the laboratory frame respectively and $X^0$ represents
the central system.
\section{A COUPLED CHANNEL ANALYSIS OF THE $K \overline K$ AND
$\pi \pi$ SYSTEMS}
\begin{figure}[h]
 \vspace{5.0cm}
\begin{center}
\includegraphics{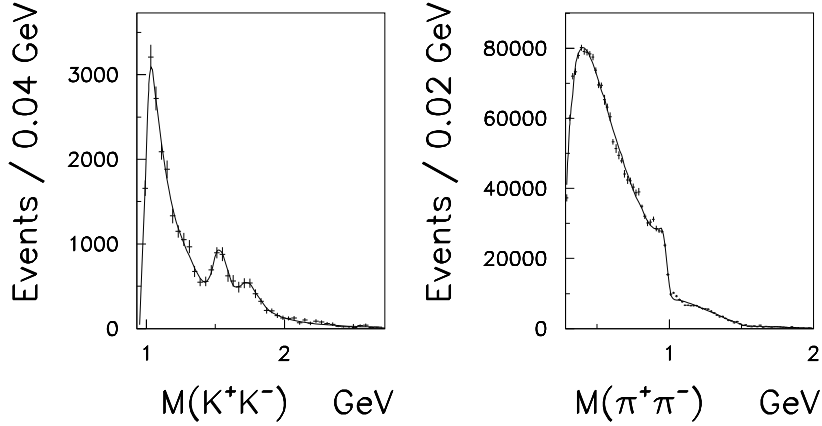}
\end{center}
\caption{ The $S_0^-$-Waves
resulting from a partial wave analysis of
the \kk  and \pipi-systems.}
\label{fi:1}
\end{figure}
\par
Recently the WA102 collaboration has published the results of
a partial wave analysis of the centrally produced \kk~\cite{re:kkpap}
and \pipi~\cite{re:pipipap} channels.
Fig.~\ref{fi:1} shows the
the $S_0^-$-wave from the \kk and \pipi channels.
The $S_0^-$-wave from the \kk channel
shows a threshold enhancement; the peaks at 1.5 GeV and
1.7~GeV are due to the
$f_0(1500)$ and $f_J(1710)$ with J~=~0~\cite{re:kkpap}.
In order to obtain a satisfactory fit
to the $S_0^-$ wave from the \pipi channel
from threshold to 2~GeV it has been found to be
necessary to use
Breit-Wigners to describe the $f_0(980)$, $f_0(1370)$
and $f_0(1500)$ and $f_J(1710)$~\cite{re:pipipap}.
\par
A coupled channel fit has been performed to the \kk
and \pipi S-wave distributions and the results are
shown in fig.~\ref{fi:1}.
The sheet II pole positions for the resonances
observed are
\begin{tabbing}
\=adfsfsf99ba \=Ma \= = \=12249 \=pi \=12000000 \=i22\=2400 \=pi \=1200000000
\=MeV   \kill
\>$f_0(980)$ \>M \>=\>($\;\;987$\>$\pm$\>$\;\;6  \pm \; \;6$)
\>$-i$\>($\;\;48$\>$\pm$\>12  $\pm \;\;  8$)\>MeV \\
\>$f_0(1370)$ \>M \>=\>(1312\>$\pm$\>25  $\pm$ 10)\>$-i$\>(109\>$\pm$\>22
$\pm$  15)\>MeV\\
\>$f_0(1500)$ \>M \>=\>(1502\>$\pm$\>12  $\pm$
10)\>$-i$\>($\;\;49$\>$\pm$\>$\;\;9$  $\pm \;$  8)\>MeV\\
\>$f_J(1710)$ \>M \>=\>(1727\>$\pm$\>12  $\pm$
11)\>$-i$\>($\;\;63$\>$\pm$\>$\;\;8$  $\pm \;$  9)\>MeV
\end{tabbing}
These
parameters are consistent with the PDG~\cite{PDG98} values for these
resonances.
For the $f_0(980)$ the couplings were determined to be
$g_\pi$~=~0.19~$\pm$~0.03~$\pm$~0.04
and $g_K$~= 0.40~$\pm$~0.04~$\pm$~0.04.
\par
The branching ratios for the $f_0(1370)$, $f_0(1500)$ and
$f_J(1710)$ have been calculated to be:
\[
\frac{f_0(1370) \rightarrow K \overline K}{f_0(1370) \rightarrow \pi \pi} =
0.46 \pm 0.15 \pm 0.11
\]
\[
\frac{f_0(1500) \rightarrow K \overline K}{f_0(1500) \rightarrow \pi \pi} =
0.33 \pm 0.03 \pm 0.07
\]
\[
\frac{f_J(1710) \rightarrow K \overline K}{f_J(1710) \rightarrow \pi \pi} = 5.0
 \pm 0.6 \pm 0.9
\]
these values are to be compared with the
PDG~\cite{PDG98} values of 1.35~$\pm$~0.68 for the $f_0(1370)$,
0.19~$\pm$~0.07 for the $f_0(1500)$, which comes from the Crystal Barrel
experiment~\cite{cbf1500}, and 2.56~$\pm$~0.9 for the $f_J(1710)$
which comes from the
WA76 experiment~\cite{oldkkpipi}.
\section{A GLUEBALL-$q \overline q$ FILTER IN CENTRAL PRODUCTION ?}
The WA102 experiment studies mesons produced in double exchange processes.
However, even in the case of pure DPE
the exchanged particles still have to couple to a final state meson.
The coupling of the two exchanged particles can either be by gluon exchange
or quark exchange. Assuming the Pomeron
is a colour singlet gluonic system if
a gluon is exchanged then a gluonic state is produced, whereas if a
quark is exchanged then a $q \overline q $ state is produced~\cite{closeak}.
In order to describe the data in terms of a physical model,
Close and Kirk~\cite{closeak},
have proposed that the data be analysed
in terms of the difference in transverse momentum ($dP_T$)
between the particles exchanged from the
fast and slow vertices.
The idea being that
for small differences in transverse momentum between the two
exchanged particles
an enhancement in the production of glueballs
relative to $q \overline q$ states may occur.
\par
The ratio of the number of events
for $dP_T$ $<$ 0.2 GeV to
the number of events
for $dP_T$ $>$ 0.5 GeV for each resonance considered has been
calculated~\cite{memoriam}.
It has been observed that all the undisputed $q \overline q$ states
which can be produced in DPE, namely those with positive G parity and $I=0$,
have a very small value for this ratio ($\leq 0.1$).
Some of the states with $I=1$ or G parity negative,
which can not be produced by DPE,
have a slightly higher value ($\approx 0.25$).
However, all of these states are suppressed relative to the
the glueball candidates the
$f_0(1500)$, $f_J(1710)$, and $f_2(1930)$,
together with the enigmatic $f_0(980)$,
which have
a large value for this ratio~\cite{memoriam}.
\begin{figure}
 \vspace{9.0cm}
\begin{center}
\includegraphics{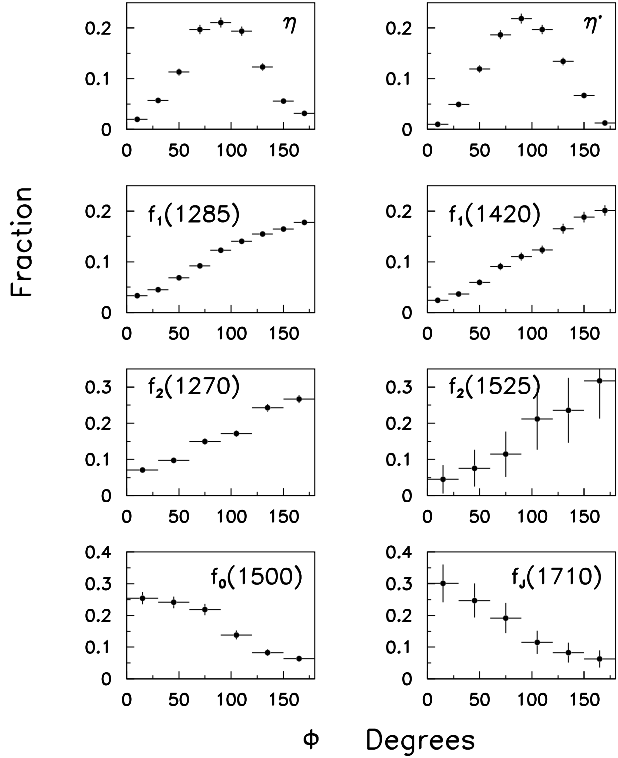}
\end{center}
\caption{The azimuthal angle between the fast and
slow protons $\phi$ for various resonances.
}
\label{fi:phidep}
\end{figure}

\section{THE AZIMUTHAL ANGLE BETWEEN THE OUTGOING PROTONS}

\par
The azimuthal angle $\phi$ is defined as the angle between the $p_T$
vectors of the two protons.
Naively it may be expected that this angle would be flat irrespective
of the resonances produced.
Fig.~\ref{fi:phidep}  shows the $\phi$ dependence for two resonances with
$J^{PC}$~=~$0^{-+}$ (the $\eta$ and $\eta^\prime$),
two with $J^{PC}$~=~$1^{++}$ (the $f_1(1285)$ and $f_1(1420)$),
two with $J^{PC}$~=~$2^{++}$ (the $f_2(1270)$ and $f_2^\prime(1525)$)
and two with $J^{PC}$~=~$0^{++}$ (the $f_0(1500)$ and $f_J(1710)$).
The $\phi$ dependence is clearly not flat and considerable variation
is observed between resonances with different $J^{PC}$s.
\par
\begin{figure}
 \vspace{5.0cm}
\begin{center}
\includegraphics{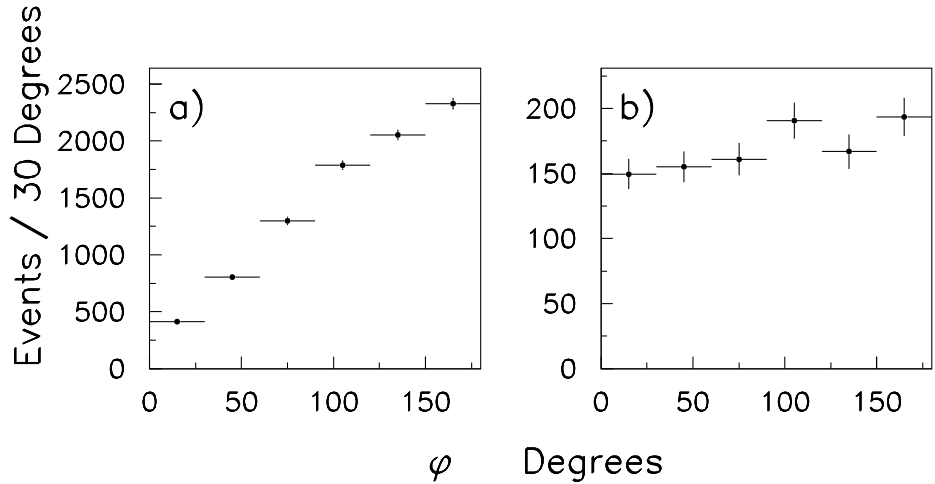}
\end{center}
\caption{The azimuthal angle between the fast and
slow protons  $\phi$  for the $f_1(1285)$ for a) $| t_1 - t_2 |$ $\le$ 0.2
and b) $|t_1 - t_2|$ $\ge$ 0.4
}
\label{fi:tdep}
\end{figure}

Several theoretical papers have been published on these
effects~\cite{angdist,clschul}.
All agree that the exchanged particle
(Pomeron) must have J~$>$~0
and that J~=~1
is the simplest explanation.
Using $\gamma^* \gamma^*$ collisions as an analogy
Close and Schuler\cite{clschul} have calculated the $\phi$ dependencies
for the production of resonances with different $J^{PC}$s.
They have found that for a $J^{PC}$~=~$0^{-+}$ state
\begin{equation}
\frac{d^3\sigma}{d\phi dt_1dt_2} \propto t_1 t_2 sin^2 \phi
\end{equation}
as can be seen from fig.~\ref{fi:phidep}
the $\phi$ distributions are proportional to $sin^2 \phi $ and
it has been found
experimentally
that $d\sigma/dt$ is proportional
to $t$ ~\cite{0mppap}.
For the  $J^{PC}$~=~$1^{++}$ states this model predicts that
$J_Z$~=~$\pm1$ should dominate, which has been found to be
correct~\cite{f1pap}, and
\begin{equation}
\frac{d^3\sigma}{d\phi dt_1dt_2}\propto (\sqrt{t_2} - \sqrt{t_1})^2 + \sqrt{t_1
t_2}sin^2 \phi/2
\label{1pp}
\end{equation}
As can be seen from fig.~\ref{fi:phidep}
the $\phi$ distributions are proportional to
$\alpha + \beta sin^2 \phi/2 $. In addition
equation(\ref{1pp})
would predict that when $|t_2 - t_1|$ is small
$d\sigma/d\phi$ should be proportional to $sin^2 \phi/2 $ while
when
$|t_2 - t_1|$ is large
$d\sigma/d\phi$ should be constant. As shown in fig.~\ref{fi:tdep} this
trend is observed in the data.
The aim now is to study the $\phi$ dependences of other
known $q \overline q$ states in order to understand
more about the nature of the Pomeron and then to use this
information as a probe for non-$q \overline q$ states.
\section{SUMMARY}
\par
In conclusion, a partial wave analysis of the centrally
produced \kk and \pipi systems has been performed.
The striking feature is the
observation of the $f_J(1710)$ with J~=~0 in
the $S_0^-$-wave.
\par
A study of centrally produced pp interactions
show that there is the possibility of a
glueball-$q \overline q$ filter mechanism ($dP_T$).
All the
undisputed $q \overline q $ states are observed to be suppressed
at small $dP_T$, but the glueball candidates
$f_0(1500)$, $f_J(1710)$, and $f_2(1930)$ ,
together with the enigmatic $f_0(980)$,
survive.
In addition, the production cross section for different
resonances depends strongly on the azimuthal angle between the
two outgoing protons which may give information on the nature
of the Pomeron.

\end{document}